# Investigation of nitrogen polar p-type doped GaN/$Al_xGa_{(1-x)}N$ superlattices for applications in wide-bandgap p-type field effect transistors

A.Krishna,[1,a)] A.Raj,[1] N.Hatui,[1] S.Keller,[1] and U.K.Mishra[1]

[1]*Department of Electrical and Computer Engineering, University of California, Santa Barbara, Goleta, California, 93106, USA*

In this study the MOCVD growth and electrical properties of N-polar modulation doped p-AlGaN/GaN superlattices (SLs) were investigated. Hole sheet charge density and mobility were studied as a function of the concentration of the p-type dopant Mg in the SL and the number of SL periods. Room temperature Hall measurements were carried out to determine the hole mobility and the sheet charge density. While the hole density increased with increasing number of SL periods, the hole mobility was largely unaffected. Hole mobilities as high as 18 $cm^2/Vs$ at a simultaneous high hole density of 6.5 x $10^{13}$ $cm^{-2}$ were observed for N-polar SLs with a Mg modulation doping of 7.5 x $10^{18}$ $cm^{-3}$. For comparable uniformly doped Ga-polar SL samples, a mobility of 11 $cm^2/Vs$ was measured. These results confirm the presence of abrupt Mg doping profiles in N-polar p-type GaN/$Al_xGa_{(1-x)}N$ SL allowing the demonstration of SLs with properties comparable to those of state-of–the-art Ga-polar modulation doped AlGaN/GaN SLs grown using MBE. Lowest sheet resistance in the GaN/AlGaN materials system of 5kΩ/□ is also reported. Test-structure transistors were also fabricated to investigate the applicability of these SL structures, with planar device resulting in a current of 5mA/mm, and a FinFET structure resulting in a current of over 100mA/mm.

Wide-Bandgap (Al,Ga)N electronic devices are at the forefront of realizing the needs of next generation communication systems, power conversion and energy conservation, thus enabling compact and affordable electronic systems[1-6]. The main reasons for this include: (a) Wide-bandgap of GaN (~3.4eV) and III-nitride alloys, enabling higher breakdown voltages[7] and (b) the existence of built-in polarization fields leading to very high mobility and charge in two dimensional-electron gas (2DEG) structures[8]. But, to tap into the full potential of this materials system, the fabrication of p-type GaN based devices is desired[9]. Some of the challenges associated with p-type GaN include: – (a) the low mobility of holes in GaN, with a maximum observed value of 40 $cm^2/Vs$ at a hole concentration of $2\times10^{12}$ $cm^{-2}$ in a two dimensional-hole gas (2DHG), and 20 $cm^2/Vs$ in bulk p-GaN (p= $1\times10^{17}$ $cm^{-3}$) [10,11]; (b) the deep nature of the common acceptor dopant in GaN (160 - 220 meV)[12,13]; (c) the passivation of MOCVD grown p-GaN material, like the one in this report, by hydrogen in the as-grown state, and the need for annealing at a high temperature (here, $800^0$C) for activation[14]; (d) the challenges involved with making ohmic contacts to p-type GaN because of the high p-GaN work function at typical doping levels[15]. Due to the above listed challenges, in the past p-channel GaN devices have received significantly less attention compared to GaN-based n-channel field effect transistors (FETs)[10,16-24].

[a)] Author to whom correspondence should be addressed. Electronic mail: athith@ucsb.edu

As holes in GaN intrinsically suffer from a low mobility (~10 $cm^2/Vs$) compared to that of electrons (1000-1500 $cm^2/Vs$)[7], a much larger number of holes is required to achieve similar currents in p-type as in n-type FETs. GaN/AlGaN superlattices (SL) are an attractive approach towards this goal. The use of Mg doped GaN/AlGaN SL is a pathway to obtain high hole concentration and mobility simultaneously. In a GaN/AlGaN SL, polarization effects create a periodic oscillation of energy bands, enhancing the ionization of deep acceptors[25]. Free carriers are separated into parallel sheets, but their spatially averaged density is much higher than in a bulk film. Mobility of holes in the 2DHG is greater than in the bulk[26]. Modulation doping the SL structure involves doping in such a way that the Mg doping is only provided away from the regions where the 2DHG forms, and hole mobilities as high as 19 $cm^2/Vs$ at a hole concentration of 1.9 x $10^{18}$ $cm^{-3}$ were measured for SL samples grown by molecular beam epitaxy[27]. The hole mobility in similar samples grown by MOCVD, however, was only about 10 $cm^2/Vs$ because of the well-known Mg surface riding effect in MOCVD preventing the formation of an abrupt doping profile[28]. All the above observations were made for Ga-polar materials grown in the typical c- or (0001) growth direction. In contrast, significantly sharper Mg doping profiles were previously observed for N-polar samples grown in the (000-1) direction[29,30]. The opposite direction of the internal electric field in N-polar compared to Ga-polar group-III nitride heterostructures and their different surface properties are attractive for enhancement mode and highly scaled transistors as well as solar cells and sensors[31].

All samples reported here were grown by MOCVD using trimethyl gallium, trimethyl aluminum, cyclopentadienyl magnesium ($Cp_2Mg$), and $NH_3$ as precursors. The SL were grown on semi-insulating (SI) N-polar GaN base layers deposited on c-plane sapphire substrates with a misorientation of $4^0$. The modulation doped SLs were composed of 4 to 10 periods, with each period being composed of 4 nm u.i.d. $Al_{0.2}Ga_{0.8}N$ + 4 nm $Al_{0.2}Ga_{0.8}N$:Mg /4 nm GaN:Mg + 4 nm u.i.d. GaN as shown in Fig.1. The first 8 nm thick p+ - GaN layer (GaN:Mg ~ 6 x $10^{19}$ $cm^{-3}$) above the SI buffer was grown to stabilize Mg flow and to achieve a sharp Mg doping profile in the epitaxial structure. The 20nm thick p+ GaN contact layer was doped with 4.5 x $10^{19}$ $cm^{-3}$ of Mg to facilitate fabrication of good contacts to the p-type SL, and was derived from work by Lund, et al[32]. Selected samples with uniform continuous Mg doping in the SL were also grown for comparison. The p-type SL stacks were grown at a temperature of $1155^0$C and pressure of 100 torr. The samples were characterized using atomic force microscopy (AFM), secondary ion mass spectroscopy (SIMS), and X-Ray diffraction (XRD).

The 2θ-ω scan of the 10 period SL sample taken around the (0004) GaN diffraction peak depicted in Fig.2. illustrates the good crystal quality and periodicity of the p-AlGaN/GaN SL.

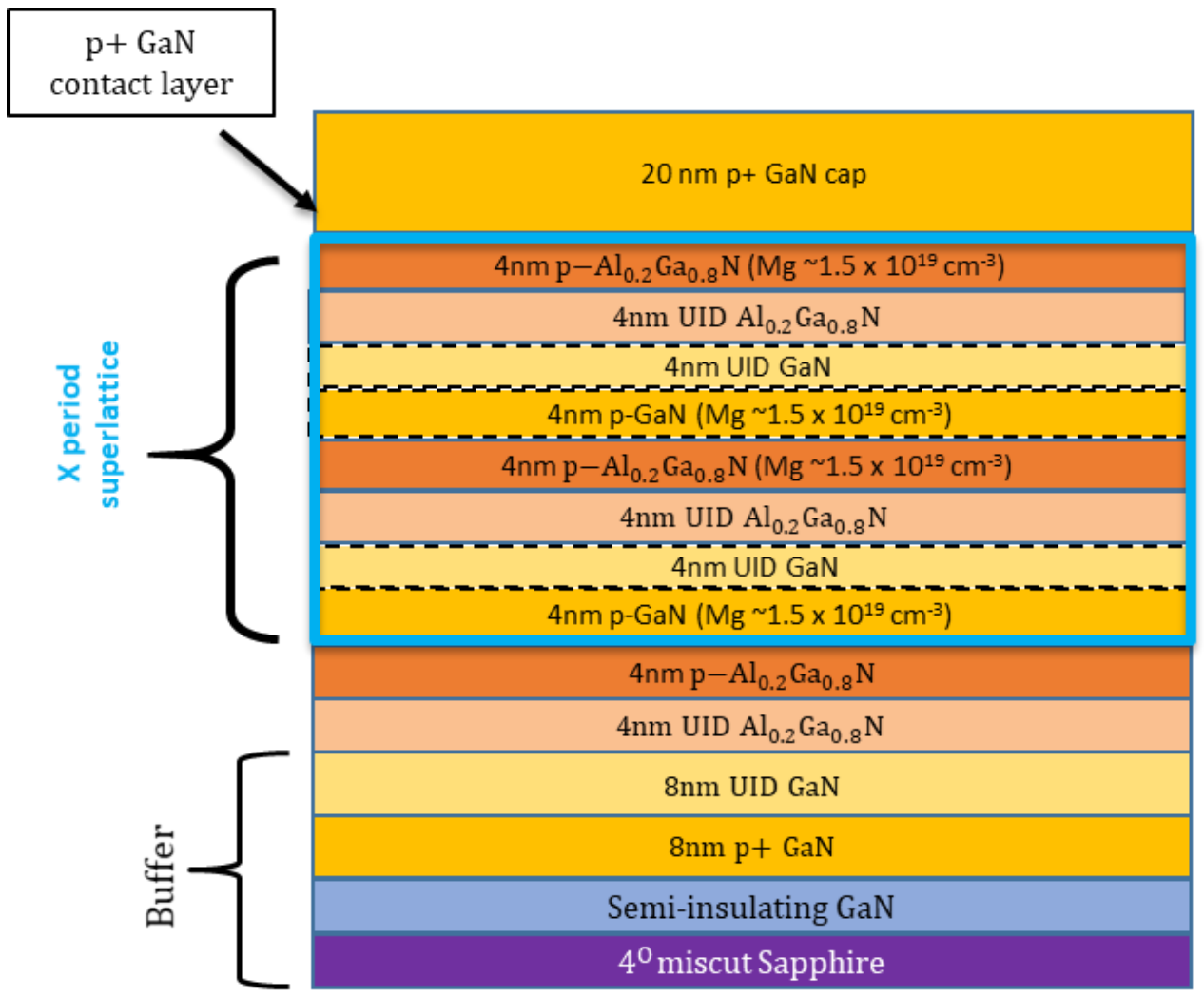

FIG. 1. Epitaxial structure of the grown N-polar modulation doped GaN/AlGaN superlattice. Samples with different number of superlattice periods and Mg doping were grown. The p+ GaN contact layer was doped with 7.5 x $10^{19}$ $cm^{-3}$ of Mg.

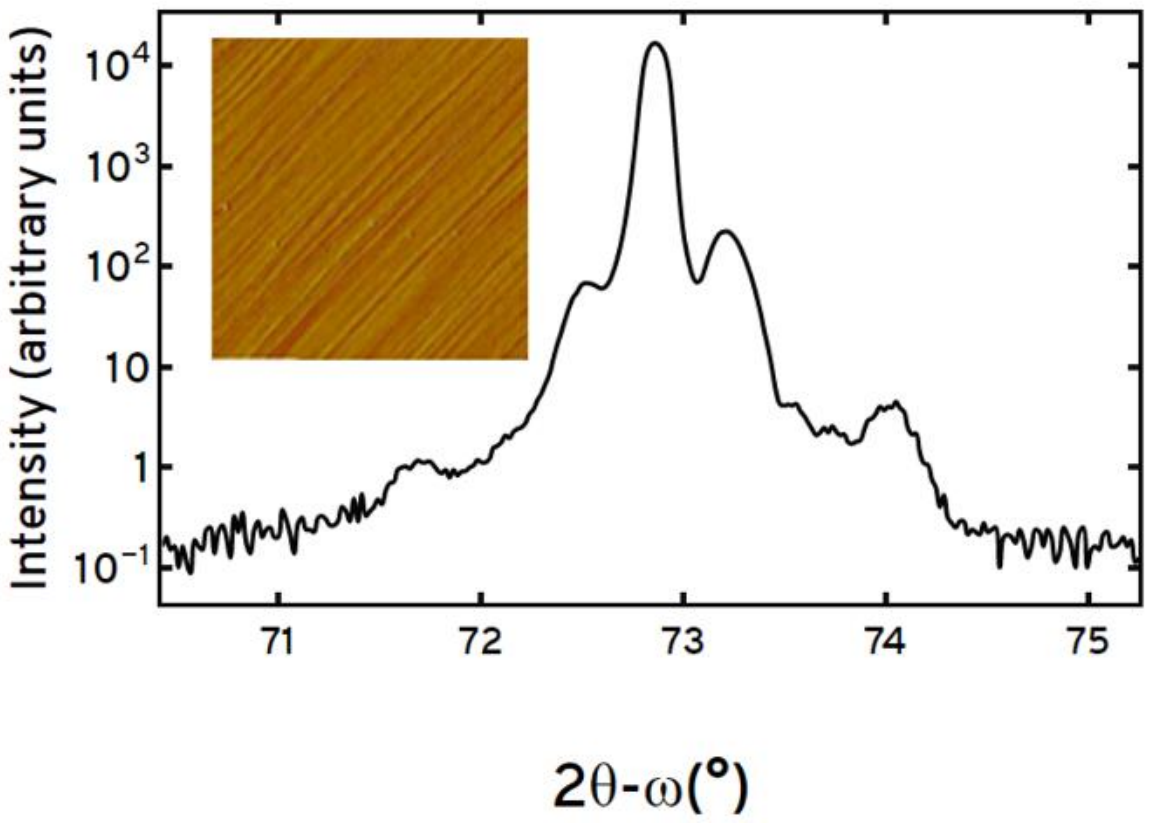

FIG. 2. 2θ-ω scan of a 10 period SL sample taken around the (0004) GaN diffraction peak. The insert displays the 10 x 10 μ$m^2$ AFM scan of the same sample (z-range = 2 nm)

Fig.3 shows the SL valence band. 8nm GaN/ 8nm AlGaN thickness was chosen as the optimum thickness from prior work done by P. Kozodoy, et. al[27]. According to this schematic, if the sample was uniformly doped, region A has a high ionization rate but low hole concentration, while region B has low ionization rate but high hole concentration. In the modulation doping scheme, only region A away from the 2DHG is doped with Mg which leads to less scattering for 2DHG from the ionized impurities resulting in a higher mobility. In MOCVD, the Mg profile for Ga-polar samples is not abrupt as seen in the work by Fichtenbaum, et al., for example, due to the well-known effect that- the Mg dopant atoms are riding on the Ga-polar surface[29]. As also reported, the Mg doping profiles in N-polar GaN are abrupt, enabling true modulation doping scheme[30].

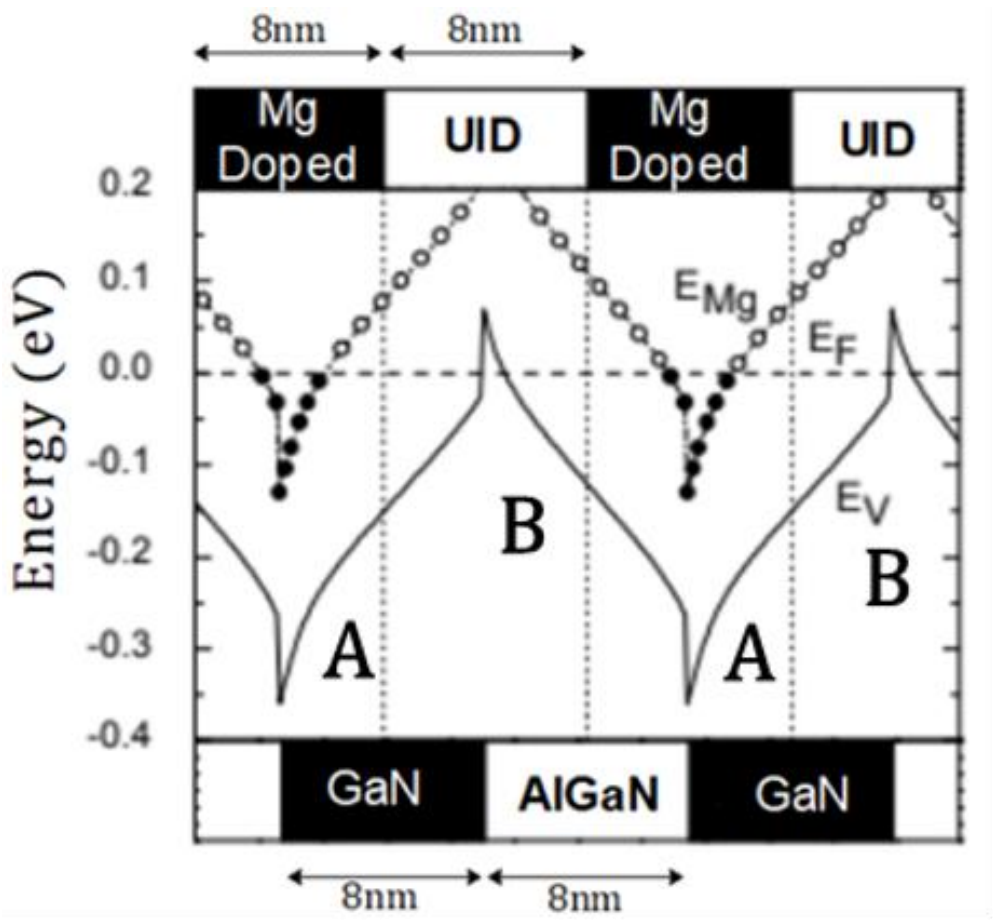

FIG. 3. Valence Band and acceptor level in the modulation doped AlGaN/GaN SL structure.

Four-point Hall measurements were carried out for the electrical characterization. Ni/Au metal stack was deposited onto 1.5 cm x 1.5 cm samples. Mg dopant activation was carried out at $800^0$C in $N_2$ environment for 5 minutes with 100 nm of $SiO_2$ deposited using PECVD, to drive the hydrogen out of the sample[24]. Fig.4 shows the Hall results for the samples - as a function of superlattice periods. The hole sheet charge increased from $3\times10^{13}$ to $6\times10^{13}$ $cm^{-2}$ as the number of superlattice periods was increased from 4 to 10 while the hole mobility remained fairly constant around 14.5 $cm^2$/Vs (average). The reduced increase in sheet charge when increasing the number SL periods from 7 to 10 was most likely associated with contacting problems. While vertical hole transport in p-SLs was reported previously[33,34], the SLs in this study were designed for lateral not vertical transport possibly limiting the number of SL periods which could be contacted. Note that a similar trend was observed for complimentary Ga-polar p-AlGaN/GaN samples, which will be reported in a separate publication[24].

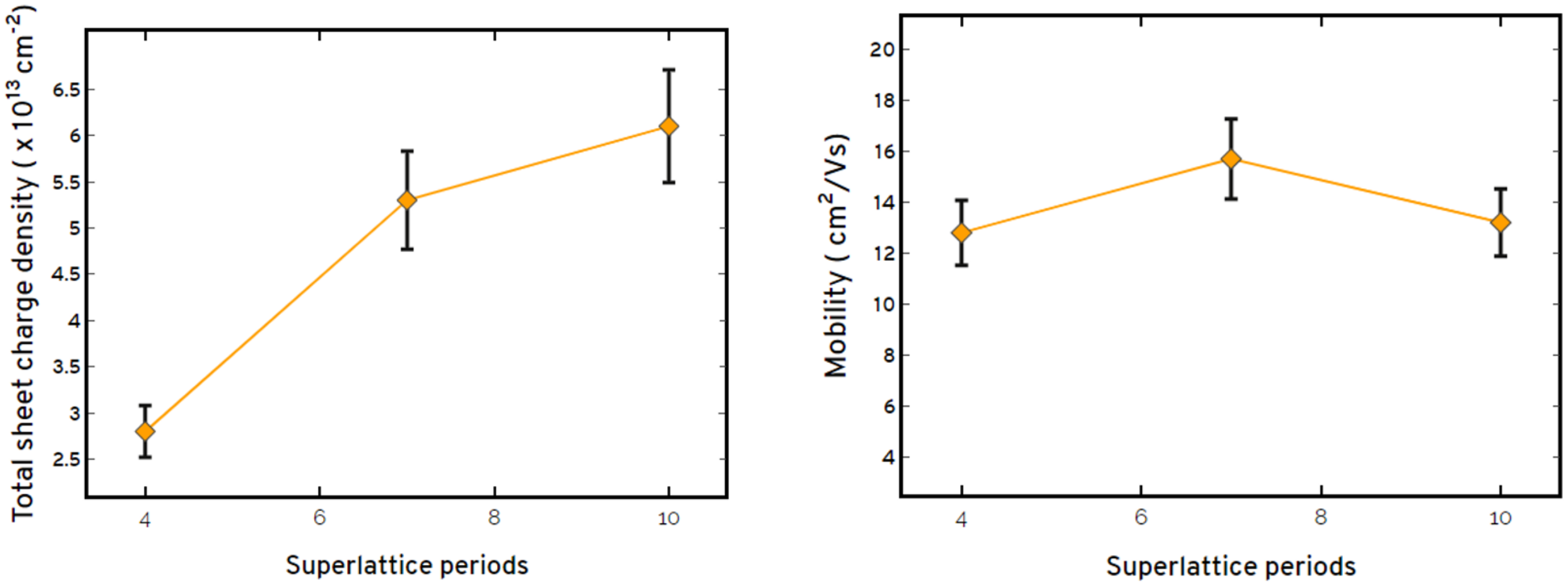

FIG. 4. (left) Total sheet charge density and (right) hole mobility in the grown structure as a function of the SL periods in samples with a Mg doping 3 x $10^{19}$ $cm^{-3}$.

Fig.5 shows the Hall results for the samples with 10 SL periods - as a function of Mg doping. The hole mobility decreased from 18.5 to 5 $cm^2/Vs$ as the doping was increased from $0.75\times10^{19}$ to $6\times10^{19}$ $cm^{-3}$. For a high sheet charge of $6\times10^{13}$ $cm^{-2}$, a hole mobility of over 18 $cm^2/Vs$ was achieved. These values are comparable to values of mobility achieved in modulation doped samples grown using MBE growth technique[28]. A uniformly doped N-polar sample was also grown for comparison. The hole sheet charge in this sample was $4.5\times10^{13}$ $cm^{-2}$ with a mobility of 6.5 $cm^2/Vs$. The reduced sheet charge compared to the modulation doped sample corresponded to an overall higher Mg concentration in the sample (Fig.5), and higher Mg content resulted in a reduced charge for all samples in this study, as can be seen in Fig. 5. The significantly lower hole mobility was a result of the increased scattering due to the presence of Mg in the 2DHG region of the SL. Similarly, the hole mobility of the best N-polar modulation doped SL sample is twice as high as that of the mobility of Ga-polar modulation doped SL samples. Note that for Ga-polar samples, no difference in the mobility between continuously and modulation doped samples was observed due to the Mg riding effect discussed before. The best N-polar modulation doped SL sample in this study exhibited the lowest sheet resistance reported amongst all reported p-channel III-nitride systems which were used for transistor applications, as can be seen in Fig.6.

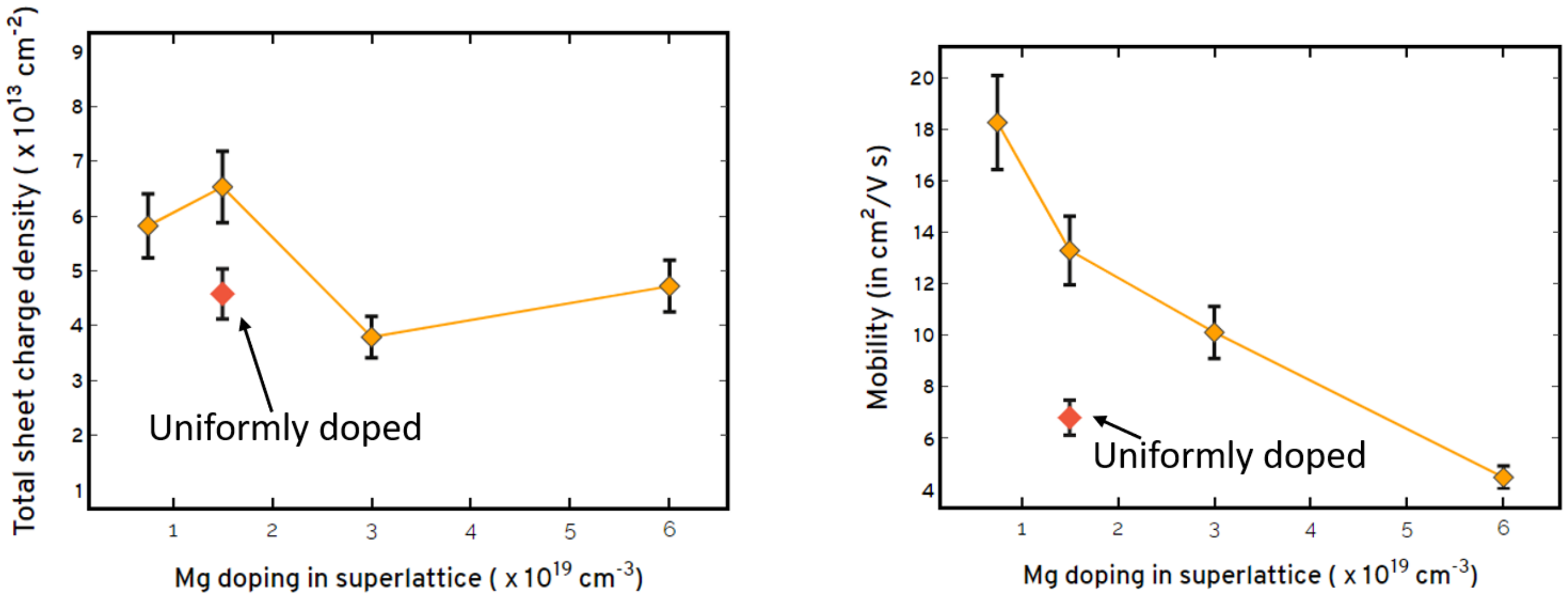

FIG. 5. (left) Total sheet charge density and (right) hole mobility in 10 period AlGaN/GaN SL samples a function of Mg doping in the SL.

Two preliminary test devices were fabricated from the SL samples :– (a) a N-polar planar p-FET with recessed SL with one p-channel (Fig.7 (a)), and (b) a N-polar SL FinFET (Fig.7(b)), to investigate the applicability of the grown SL samples for field effect transistors (FET). A Ni/Au stack was used for the source and drain contacts, a Ti/Au stack was used for the gate contacts and $SiO_2$ was used as the gate dielectric for both the planar and FinFET device. The planar recessed p-FET was fabricated from a 7 period SL sample with a Mg doping of 1.5 x $10^{19}$ $cm^{-3}$ and exhibited a maximum drain-source current of 5.5 mA/mm, and an On-resistance of 1.82 kΩ.mm. The device had a turn on voltage of 2.5V due to the lack of Ohmic contacts to all SL channels. To use all the SL channels, the FinFET was fabricated with the gate wrapped around the 10 SL channels in a sample with Mg doping of 0.75 x $10^{19}$ $cm^{-3}$. This device resulted in maximum drain-source current of 120 mA/mm with the fin-width being 300nm. The device dimensions for both the planar and FinFET device are given in Table 1. Fig.8(a) shows the Current-Voltage characteristics for the planar device, and Fig.8(b) that of the FinFET. Presently, experiments are being carried out to eliminate the turn-on in the devices, while device scaling experiments are being carried out to pinch-off the FinFET.

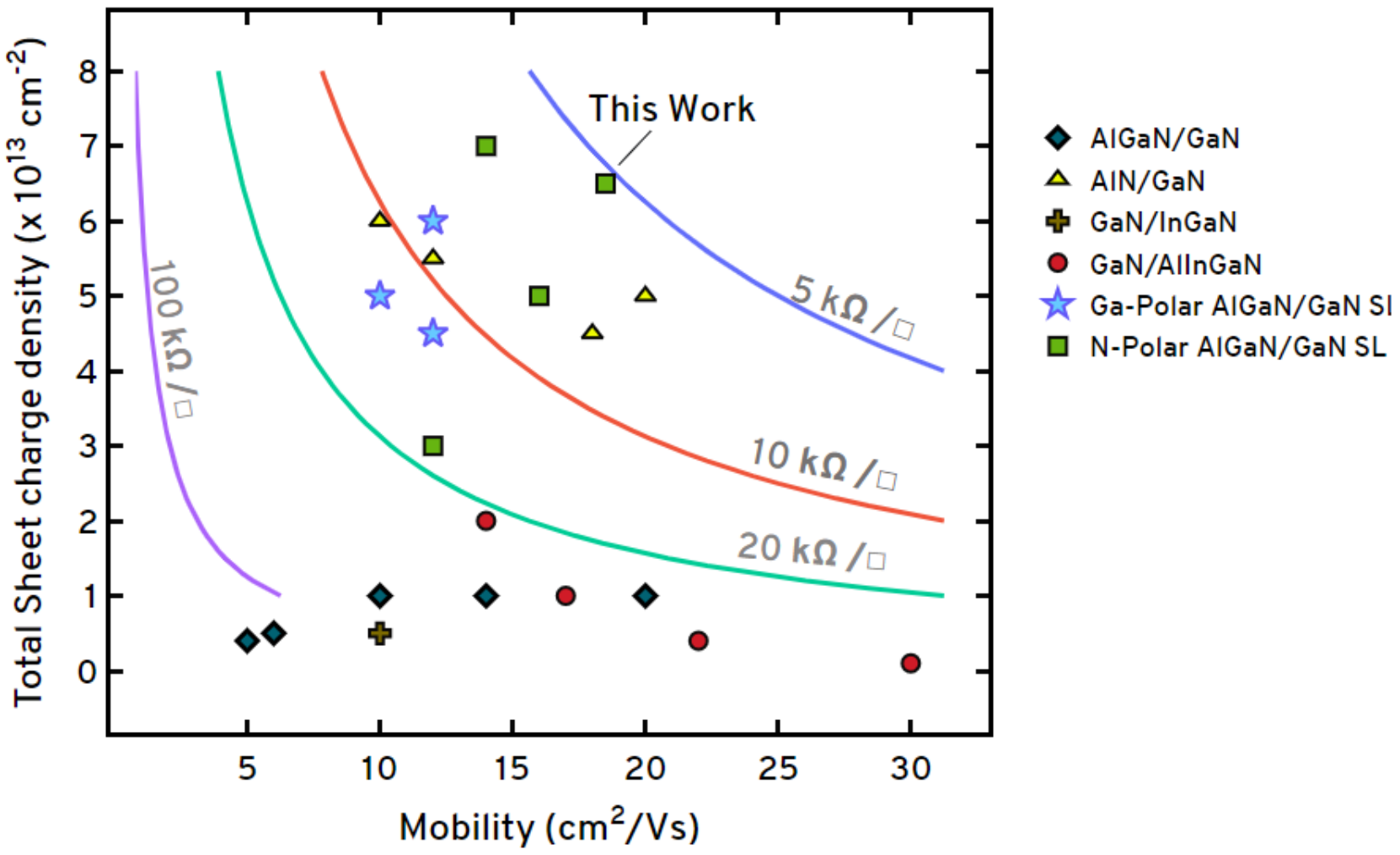

FIG. 6. Sheet charge density and mobility data for the reported p-channel III-nitride systems in literature which were utilized for transistor applications. Our work, N-polar modulation doped AlGaN/GaN SL has the lowest sheet resistance reported using MOCVD growth technique[10,16-24].

In conclusion, N-Polar Modulation doped GaN/AlGaN SLs were grown with high sheet charge of 6 x $10^{13}$ $cm^{-2}$ and mobility of over 18 $cm^2$/Vs resulting in a sheet resistance of ~ 5kΩ/□, lowest amongst all p-type III-nitride systems. In comparison, the Ga-polar samples had sheet charge of 4 – 6 x $10^{13}$ $cm^{-2}$ and mobility of 10 $cm^2$/Vs[24]. Applicability of these SL structures was tested using a planar device and a FinFET showing promising results. Further improvements in the device performance are expected with further process optimization including scaling of the fins in the FinFET, as well as N-polar processing optimization.

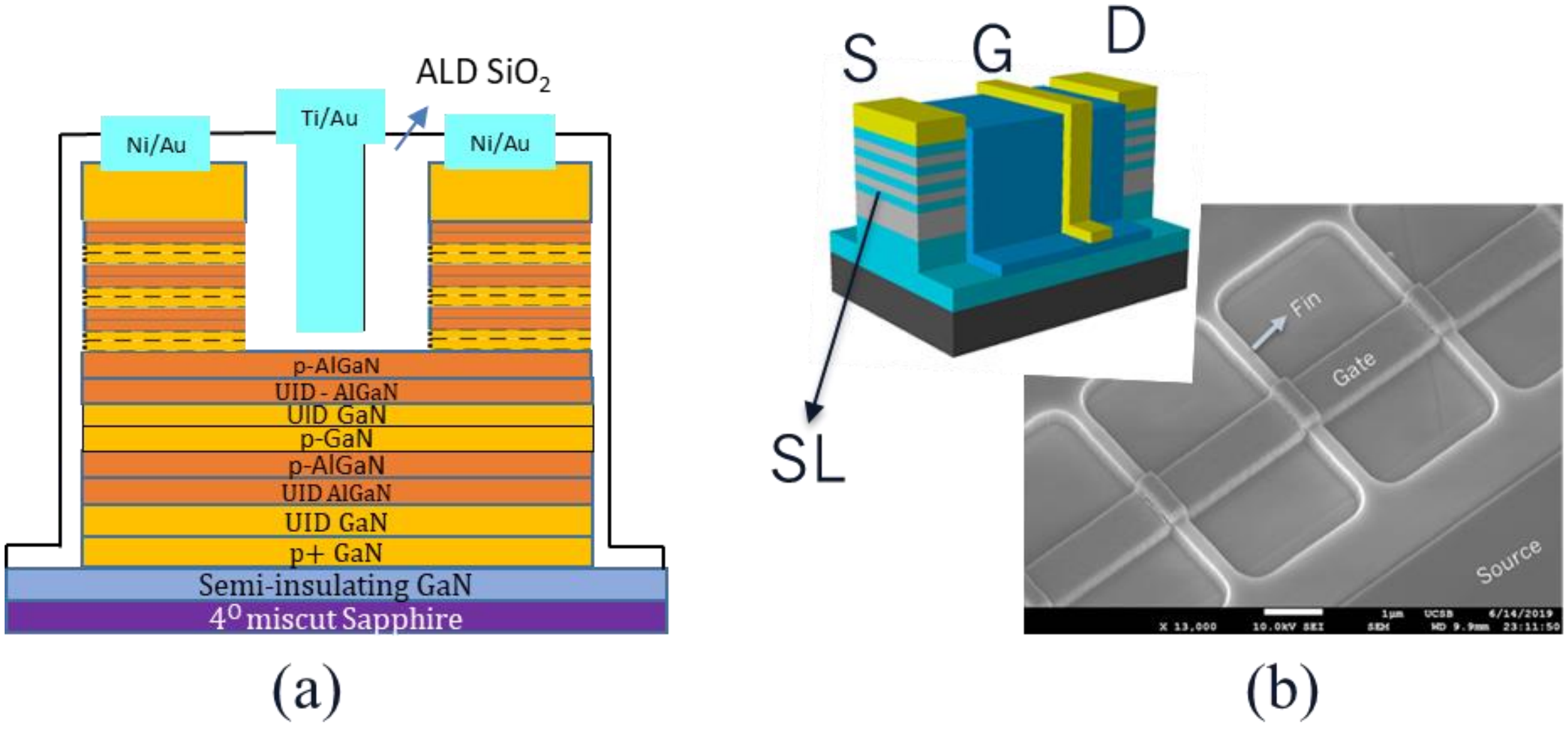

FIG. 7. (left) (a) Schematic of N-polar Planar, SL recessed FET; (right) (b) Schematic of N-polar SL FinFET with the SEM image of the processed FinFET device

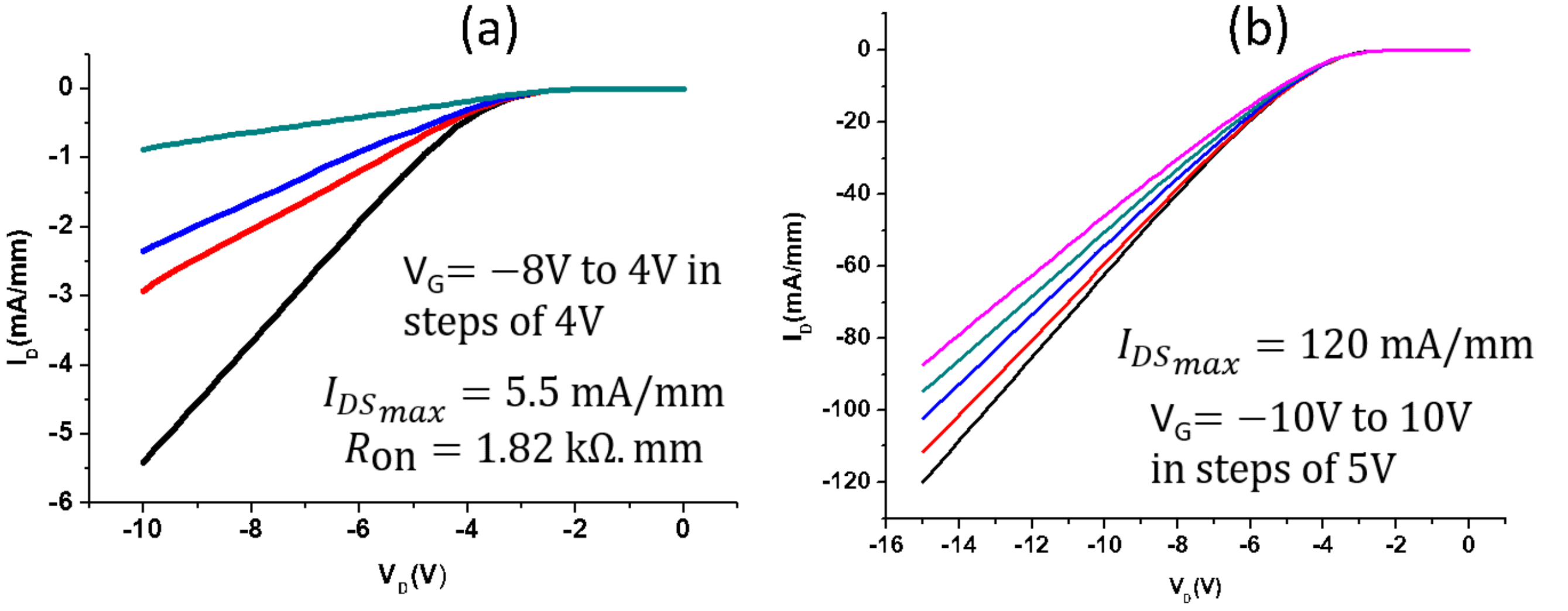

FIG. 8. (left) (a) Electrical characterization of the N-polar Planar, SL recessed FET; (right) (b) Electrical characterization of the N-polar SL FinFET

TABLE I. Device dimensions for the N-polar planar and FinFET device.

| | (a) N-polar Planar, SL recessed FET | (b) N-polar SL FinFET |
|---|---|---|
| Gate Length | 0.7 µm | 1 µm |
| Gate Source spacing | 0.4 µm | 3 µm |
| Gate Drain spacing | 0.4 µm | 4 µm |

This work was supported by ASCENT, one of six centers in JUMP, a Semiconductor Research Corporation (SRC) program sponsored by DARPA.